\documentclass[12pt]{article}
\def\beq{\begin{equation}}
\def\eeq{\end{equation}}
\def\pp{\psi'}
\def\ppp{\psi(3770)}
\def\gsim{\raisebox{-0.6ex}{$\stackrel{\textstyle >}{\sim}$}}

\begin{document}
\begin{titlepage}
\begin{center}
{\Large \bf William I. Fine Theoretical Physics Institute \\
University of Minnesota \\}  \end{center}
\vspace{0.2in}
\begin{flushright}
FTPI-MINN-05/12 \\
UMN-TH-2354-05 \\
April 2005 \\
\end{flushright}
\vspace{0.3in}
\begin{center}
{\Large \bf  The $\bar c c$ purity of $\psi(3770)$ and $\psi'$ challenged
\\}
\vspace{0.2in}
{\bf M.B. Voloshin  \\ }
William I. Fine Theoretical Physics Institute, University of
Minnesota,\\ Minneapolis, MN 55455 \\
and \\
Institute of Theoretical and Experimental Physics, Moscow, 117259
\\[0.2in]
\end{center}

\begin{abstract}

It is suggested that the resonance $\psi(3770)$ may contain a sizeable
($O(10\%)$ in terms of the probability weight factor) four-quark component with
the up- and down- quarks and antiquarks in addition to the $c {\bar c}$ pair,
which component in itself has a substantial part with the isospin $I=1$.
Furthermore such four-quark part of the wave function should also affect the
properties of the $\psi'$ charmonium resonance through the $\psi(3770) - \psi'$
mixing previously considered in the literature. It is argued that an admixture
of extra light quark pairs can explain a possible discrepancy between the
theoretical expectations and the recent data on the non-$D {\bar D}$ decay width
of the $\psi(3770)$ and the ratio of the yield of charged and neutral $D$ meson
pairs in its decays, as well as on the extra rate of the $\psi'$ direct decay
into light hadrons and the rate of the decay $\psi' \to \pi^0 \, J/\psi$. It is
further argued that the suggested four-quark component of the wave function of
the $\psi(3770)$ should give rise to a measurable rate of the decays $\psi(3770)
\to \eta \, J/\psi$ and $\psi(3770) \to \pi^0 \, J/\psi$.

\end{abstract}

\end{titlepage}

\section{Introduction}
The study of charmonium resonances below and above the $D {\bar D}$ threshold
currently experiences a remarkable revival due to the efforts of the CLEO-c
experiment\cite{cleoc}. While the general picture of the properties of these
resonances is certainly in agreement with the theoretical expectations for
states of $c {\bar c}$ charmonium system, the fine details emerging with the
improvement in the accuracy of the data on these resonances possibly bring into
light some yet unsolved issues of the charmonium physics, as well as of a
general understanding of the QCD dynamics. The purpose of this paper is to point
out four specific pieces of the recent experimental data on the $\pp$ and $\ppp$
resonances, which are potentially at variance with the standard theoretical
expectations, and which can possibly be explained by assuming a certain
admixture of a four-quark component with extra $u {\bar u}$ and $d {\bar d}$
quark pairs. Namely, the data to be discussed are:\\
the total cross section for the $D {\bar D}$ pair production in $e^+e^-$
annihilation at the $\ppp$ peak\cite{cleo1}
\beq
\sigma(e^+e^- \to D {\bar D}) = (6.39 \pm 0.10 ^{+0.17}_{-0.08})\, {\rm nb};
{\label{data}}
\eeq
the charged to neutral yield ratio at the same energy\cite{cleo1}
\beq
\sigma(e^+e^- \to D^+  D^-)/\sigma(e^+e^- \to D^0 {\bar D}^0)=0.776 \pm
0.024^{+0.014}_{-0.006};
\label{datb}
\eeq
the branching fraction for the direct decays of the $\pp$ resonance into light
hadrons\cite{cleo2}
\beq
{\cal B}(\pp \to {\rm light~hadrons})=(16.9 \pm 2.9)\, \%;
\label{datc}
\eeq
and the newly measured\cite{cleo2} ratio of the branching fractions
\beq
{\cal B}(\pp \to \pi^0 J/\psi)/{\cal B}(\pp \to \eta J/\psi)=(4.1 \pm 0.4 \pm
0.1) \, \%.
\label{datd}
\eeq

In what follows the discrepancy of each of these experimental numbers with
previous theoretical expectations in the picture where both the $\pp$ and $\ppp$
resonances are pure $c {\bar c}$ states is discussed in Sect.2, and, where
possible, alternative explanations of the discrepancy within the same picture
are mentioned. It will be then argued in Sect.3 that the listed experimental
data can be explained, at least semi-quantitatively, by assuming the presence of
light quark pairs in a part of the wave function of the $\ppp$, and also in a
much lesser part of the wave function of the $\pp$, where the latter may arise
from a $\pp - \ppp$ mixing of the type considered by Rosner\cite{rosner}, as
discussed in Sect.4. The presence of extra pairs of light quarks in the
resonances arguably affects the data (\ref{data}) and (\ref{datc}) in the list
above, while for an explanation of the data (\ref{datb}) and (\ref{datd}) one
has to assume that in the suggested four-quark admixture there is a substantial
component with the isospin $I=1$, i.e. to assume an unequal weight of the states
with extra $u {\bar u}$ and $d {\bar d}$ pairs. As will be further argued in
Sect.5, both these assumptions can be tested experimentally by measuring the
decays $\ppp \to \eta J/\psi$ and $\ppp \to \pi^0 J/\psi$, of which the former
should be enhanced due to the four-quark component in $\ppp$ with $I=0$, while
the latter is expected to be enhanced due to the component with $I=1$. Finally,
the Sect.6 contains the summary and discussion of the arguments presented here
and a discussion of the current and possible future experimental data.

\section{The data vs. the theoretical expectations}
We proceed to the discussion of the data listed in eqs.(\ref{data} - \ref{datd})
starting with the last one. The ratio of the decay rates in eq.(\ref{datd}) was
considered in Ref.\cite{is} using an extension of the chiral algebra technique
from a previous work\cite{bli} and also in the context of the QCD multipole
expansion for hadronic transitions in heavy quarkonium\cite{gottfried,mv}.
Within this approach both the $\pp$ and $J/\psi$ are considered as compact $c
{\bar c}$ states, so that in their interaction with soft gluon field they are
equivalent to a point-like source. The ratio of the decay amplitudes is then
fully determined by the ratio of the amplitudes for creation of the
corresponding light pseudoscalar meson by the gluonic operator $G {\tilde G}=
(1/2)\epsilon_{\mu \nu \lambda \sigma} \, G^{\mu \nu} G^{\lambda \sigma}$. The
ratio of the latter amplitudes is determined by the anomaly in the axial current
and the isospin and the flavor SU(3) breaking by the light quark
masses\cite{aanom}.  In terms of the decay rates the result for the ratio reads
as\cite{is}
\beq
{\Gamma(\pp \to \pi^0 J/\psi)\over \Gamma(\pp \to \eta J/\psi)}=3 \,\left (
{m_d-m_u \over m_d + m_u } \right )^2 \, {f_\pi^2 \over f_\eta^2} \, {m_\pi^4
\over m_\eta^4} \, {p_\pi^3 \over p_\eta^3}~,
\label{petar}
\eeq
where $p_\pi$, $p_\eta$ is the momentum of the corresponding light meson in the
decay, $m_u$ and $m_d$ are the light quark masses, and $f_\pi$ and $f_\eta$ are
the annihilation constants of the mesons, normalized in such a way that  in the
limit of the flavor SU(3) symmetry $f_\pi=f_\eta$. In reality it is known from
comparison of $f_\pi$ and $f_K$ that the presence of heavier strange quarks
increases the constant $f$, so that $f_\eta$ is expected to be larger than
$f_\pi$. Therefore the limit $f_\pi=f_\eta$ can be used as an upper bound on the
ratio of the rates in eq.(\ref{petar}). The ratio of the masses of the $u$ and
$d$ quarks, describing the explicit breaking of the chiral symmetry and the
isospin violation in this breaking was studied years ago in great detail by
Gasser and Leutwyler\cite{gl}. The largest value for the ratio $(m_d-m_u)/(m_d +
m_u)$ allowed by that study is approximately 0.35.
Thus the theoretical upper bound for the ratio of the decay rates in
eq.(\ref{petar}) is approximately $2.3 \%$, which is still by more than
$4\sigma$ below the experimental result (\ref{datd}).
It can be also mentioned in connection with the light quark mass ratio that the
well known Weinberg's formula\cite{weinberg} gives
\beq
{m_d-m_u \over m_d + m_u }={m_{K^0}^2-m_{K^+}^2+m_{\pi^+}^2-m_{\pi^0}^2 \over
m_{\pi^0}^2} = 0.285~,
\label{weinf}
\eeq
and results in a still smaller ratio of the decay rates if used in
eq.(\ref{petar}).
It is certainly understood\cite{is} that the formula (\ref{petar}) may receive
unexpectedly large corrections from the effects of the SU(3) violation, however
such corrections can also significantly affect the analysis of the chiral
phenomenology in Ref.\cite{gl}, and the whole subject then would have to be
revisited anew. It should be mentioned that the largest theoretical estimate of
the ratio (\ref{datd}) found in the literature\cite{kty} corresponds to $3.4
\%$, which is also significantly lower than the experimental number. However,
the latter estimate does not fully take into account the proper QCD structure of
the relevant amplitude for the meson production by the gluonic operator. A
detailed consideration\cite{mv2} of the terms omitted in the earlier
papers\cite{vz,is} reinstates the formula in eq.(\ref{petar}) for the ratio of
the decay rates.

The rate of the direct decays of the $\pp$ into light hadrons, has been a source
of concern previously, and the latest experimental result (\ref{datc})
emphasizes the problem at a somewhat greater statistical significance. In the
standard picture of the three-gluon annihilation of the $^3S_1$ charmonium
states both the hadronic and the $\ell^+ \ell^-$ annihilation amplitudes are
proportional to the wave function at the origin, so that the ratio of the rates
of these processes should be the same for $J/\psi$ and $\pp$. However, the
combination of the PDG values\cite{pdg} gives
\beq
{\cal B}(\pp \to \ell^+ \ell^-) \, {{\cal B}(J/\psi \to {\rm
light~hadrons})\over {\cal B}(J/\psi \to \ell^+ \ell^-)} = (10.9 \pm 0.6)\, \%~,
\label{ppjpth}
\eeq
which is by $2.2\sigma$ below the experimental number in eq.(\ref{datc}).
Clearly, the uncertainty in the latter number is still large enough to
accommodate compatibility with the standard picture. Nevertheless, even given
the present uncertainty, it makes sense to look into corrections to the
straightforward prediction of the equality between the combination of the
branching fractions in eq.(\ref{ppjpth}) and that in eq.(\ref{datc}). Moreover,
it is well known that the similarity of the annihilation decay rates is strongly
broken in exclusive channels, most notably in the decays of $J/\psi$ and $\pp$
into $\rho \pi$. (For a recent discussion see e.g. Ref.\cite{rosner}.) One
source of modification of the short-distance hadronic annihilation of a heavy
quarkonium is provided by the non-perturbative corrections\cite{mv3} due to the
gluon condensate. These corrections depend on details of the quarkonium wave
function, which are not reduced to the wave function at the origin. For this
reason the estimate of the non-perturbative behavior can be done in a model
independent way only for a very heavy quarkonium, for which the dynamics is
dominated by a Coulomb-like short distance gluon force. An extrapolation down to
the realistic bottomonium and charmonium suggests that this effect should
somewhat suppress the hadronic annihilation of the 2S state relative to the 1S.
The recent experimental data\cite{galik} on the $\ell^+ \ell^-$ branching
fractions for the $\Upsilon$ resonances, ${\cal B}_{\mu \mu}(\Upsilon)= (2.53
\pm 0.02 \pm 0.05) \, \%$,  ${\cal B}_{\mu \mu}(\Upsilon')= (2.11 \pm 0.03 \pm
0.05) \, \%$ tend to support such behavior, since they correspond to the ratio
\beq
{{\cal B}_{\mu \mu}(\Upsilon') \, {\cal B}(\Upsilon \to {\rm light~
hadrons})\over {\cal B}_{\mu \mu}(\Upsilon) \, {\cal B}(\Upsilon' \to {\rm
light~ hadrons})} \approx 1.5 \pm 0.1~.
\label{upsr}
\eeq
If this behavior is extrapolated further down to the charmonium mass, the
non-perturbative corrections in fact enhance the disagreement between the
numbers
in eq.(\ref{datc}) and (\ref{petar}).

Proceeding to discussion of the data on the $\ppp$ resonance we first notice
that the measured charged-to-neutral yield ratio (\ref{datb}) significantly
exceeds the ratio of the $P$ wave kinematical factors in the corresponding
channels, which is equal to $p_+^3/p_0^3 =0.685$. If the $\ppp$ and the $D$
mesons were point particles, the yield of the $D^+ D^-$ pairs would be enhanced
by the well known Coulomb factor\cite{am} $R_c=1+\pi \alpha/(2 v_+) \approx
1.085$, making the expected ratio equal to 0.743 in a reasonable agreement with
the data (\ref{datb}). It is well understood however, that in a more realistic
picture taking into account the form factors of the $D$ mesons\cite{lepage} and
of the $\ppp \to D {\bar D}$ vertex\cite{be} the Coulomb enhancement is
significantly weaker. Moreover, if the $\ppp$ resonance is considered as
`strong' in the sense that the $P$ wave dynamics of the $D$ mesons at energies
close to the resonance peak is totally dominated by the Breit-Wigner form of the
wave function, the interference of the Coulomb and the resonance phase shifts
results in a nontrivial energy behavior of the charged-to-neutral yield ratio,
and generally almost completely eliminates the Coulomb enhancement at the energy
close to that of the peak\cite{mv4}. This behavior generally agrees with the
measurements by CLEO\cite{cleo1m,cleo2m}, $BABAR$\cite{babar1,babar2}
and Belle\cite{belle} at the peak energy of the $\Upsilon(4S)$ resonance, where
they find a very weak or nonexistent Coulomb enhancement of the yield in the
$B^+B^-$ channel, far below the point-like estimate 1.19 for the Coulomb factor.

Finally, as recently pointed out by Rosner\cite{rosner2,rosner3}, the measured
total cross section for the $D {\bar D}$ production in the $e^+ e^-$
annihilation at the $\ppp$ peak (eq.(\ref{data})) is lower than the previous
measurements (by other groups) of the total resonant cross section, for which he
estimates the average of $(7.9 \pm 0.6) \,$nb. The approximately $(1.5 \pm
0.7)\,$nb deficit of the cross section, if confirmed by the new data, would have
to be covered by direct decays of the $\ppp$ into light hadrons, since the
expected hadronic and radiative transitions from $\ppp$ to lower charmonium
states can account\cite{rosner3} for at most about 2\% of its total width $23.6
\pm 0.7 \,$MeV. If $\ppp$ is a pure $c {\bar c}$ state with $J^{PC}=1^{--}$, it
would be practically impossible to explain an annihilation width in excess of
few tens of KeV, and thus to understand any significant non-$D {\bar D}$ cross
section at the resonance.

\section{The four-quark component and the properties of $\ppp$}
If the discussed discrepancies between the experimental data and the theoretical
expectations based on the picture where both the $\ppp$ and $\pp$ are pure $c
{\bar c}$ states are taken seriously, it is quite likely that such picture has
to be modified. In particular Rosner considers\cite{rosner3} a model of
re-annihilation of the $D {\bar D}$ pairs in order to explain possible non-$D
{\bar D}$ decays of the $\ppp$. The re-annihilation model can be viewed as a
particular dynamical implementation of the picture suggested here that the wave
function of the $\ppp$ resonance contains at the characteristic hadronic
distances a sizeable admixture of four-quark states with the $u {\bar u}$ and $d
{\bar d}$ light quark pairs in addition to a $c {\bar c}$ pair.
In other words it is assumed that the `core' of the wave function of the $\ppp$
resonance consists of the following three essential parts:
\beq
\Psi_{\psi''}= a \, c {\bar c} + b_0 \, c {\bar c} \, (u {\bar u} + d {\bar
d})/\sqrt{2} + b_1 \, c {\bar c} \, (u {\bar u} - d {\bar d})/\sqrt{2}~,
\label{decomp}
\eeq
where $a$, $b_0$, and $b_1$ are coefficients, and the $I=1$ part proportional to
$b_1$ is written for the reasons to be discussed further. Naturally, the
expression (\ref{decomp}) is rather symbolic, since each of the terms in it
contains further specifications, such as the coordinate wave functions as well
as color combinations for the four-quark part, etc. Lacking the knowledge of
these details we are bound to limit ourselves to a discussion of only the
generic properties following from the assumed in eq.(\ref{decomp}) quark flavor
structure of the $\ppp$. The assumed presence of the four-quark states
can certainly be a result of a very strong coupling between the $D {\bar D}$
channel and the $c {\bar c}$ state. Unlike the re-annihilation scheme this
picture does not need to rely on the assumption that the $D$ mesons can be
considered as individual `intact' mesons at the distances, where they actually
overlap. On the other hand the `generic' scheme discussed here only allows to
make general semi-quantitative predictions.

It is well-known\cite{ov} that the $c {\bar c}$ pair inside a four-quark
component can annihilate in the second order in the QCD coupling $\alpha_s$,
i.e. much faster than a colorless $J^{PC}=1^{--}$ heavy quark pair, which is
bound by the conservation laws to annihilate via three gluons. In particular a
$^3S_1$ $c {\bar c}$ pair in a color octet state annihilates into light quarks,
$c {\bar c} \to q {\bar q}$ via one gluon\cite{bgr,ov2}. On the other hand it is
quite likely that a color octet  $^3S_1$ $c {\bar c}$ state is present in the
four-quark part with a weight comparable to one. Indeed the spin-triplet states
of the heavy quark pair should dominate in the Fock decomposition in
eq.(\ref{decomp}) due to the spin selection rule\cite{mv5}, since the leading
term proportional to $a$ corresponds to spin-triplet $^3S_1$ and $^3 D_1$
states. Furthermore, if in the four-quark component the relative color of the
$c$ and ${\bar c}$ is randomized, the states with the octet overall color of the
pair carry the statistical weight of 8/9. Once the annihilation of the heavy
quark pair from the four-quark component in order $\alpha_s^2$ is allowed, the
proper decay rate of this component can be deduced from a typical $\alpha_s^2$
annihilation rate of $c {\bar c}$ in an $S$ wave state, of which the only
measured example is the decay width of the $\eta_c$. Thus one arrives at an
(rather approximate) estimate
\beq
{\cal B}[\ppp \to {\rm light~ hadrons}] \sim  \left ( |b_0|^2 + |b_1|^2 \right )
\, {\Gamma(\eta_c) \over \Gamma[\ppp)]}~.
\label{gnondd}
\eeq
Experimentally, the total width of $\eta_c$ is still somewhat uncertain, but is
comparable to that of the $\ppp$. Thus an $O(10\%)$ non-$D {\bar D}$ branching
fraction in the decays of $\ppp$ requires the weight factor $( |b_0|^2 +
|b_1|^2)$ for the four-quark component in eq.(\ref{decomp}) of a similar value.
Given the uncertainty in the current data on the non-$D {\bar D}$ decays of
$\ppp$, it would be premature to quantitatively pursue this point any further.

If the assumed four-quark component in the $\ppp$ arises from a strong mixing
with the $D {\bar D}$ channel, one can expect an enhanced breaking of the
isospin in this component due to a substantial difference in the excitation
energy for the $D^+ D^-$ and $D^0 {\bar D^0}$ mesons: approximately 30 MeV vs.
40 MeV. This breaking, corresponding to a nonzero value of the coefficient $b_1$
in eq.(\ref{decomp}), can be related to the larger than expected relative yield
of the charged meson pairs $D^+ D^-$ (cf. eq.(\ref{datb})). Indeed, the
calculation\cite{mv4} resulting in essentially no Coulomb effect at the peak of
the resonance assumes a perfect isotopic symmetry in the `core' of the
resonance, i.e. in the region of strong interaction. If this assumption is
invalidated by the presence of an $I=1$ four-quark component at those distances,
the ratio of the wave functions for the channels with charged and neutral mesons
would be biased proportionally to the coefficient $b_1$, resulting in a shift of
the yield ratio by
\beq
\Delta \left [ {\sigma(e^+e^- \to D^+  D^-) \over \sigma(e^+e^- \to D^0 {\bar
D}^0)} \right ] \propto -4 \, b_1 \, \left ( {p_+   \over p_0} \right )^3~.
\label{shift}
\eeq
Although the proportionality coefficient in this relation is not known at the
present level of understanding, it looks quite reasonable to conclude that an
$O(-3\%)$ value of the coefficient $b_1$ would be sufficient for explaining an
$O(10\%)$ difference between the data (\ref{datb}) and the theoretical
predictions based on the assumption that the $\ppp$ is a pure isoscalar.

\section{The effects of the four-quark component in $\pp$}
The presence of a four-quark component in the $\ppp$ can also affect the
properties of the $\pp$ resonance through the $\ppp - \pp$ mixing. Such mixing
was recently analyzed by Rosner\cite{rosner} in the `pure' charmonium model as a
$^3S_1 - ^3D_1$ mixing. Based on the experimental values of the $\ell^+ \ell^-$
widths of the discussed resonances and on model calculations of these widths for
$^3S_1$ and $^3D_1$ charmonium, he found the mixing angle to be $(12\pm 2)^o$.
It should be understood however that this estimate of the mixing is far from
being final, not only due to its theoretical model dependence, but also because
the experimental data, especially for $\ppp$, are currently in flux and are
likely to change with further progress of the experiment. Indeed, the current
data on $\Gamma_{\ell \ell}[\ppp]$, $\Gamma_{\rm tot}[\ppp]$, and on the
$e^+e^-$ cross section at the peak are in an apparent contradiction with the
unitarity formula for the cross section at the maximum:
\beq
\sigma[e^+e^- \to \ppp]={12 \, \pi \over M^2} \, {\cal B}_{ee}~.
\label{smax}
\eeq
Using the PDG data\cite{pdg} for the branching fraction ${\cal B}_{ee}$ and the
mass $M$ of the $\ppp$ one finds from this constraint the value $(11.9 \pm
1.8)\,$nb for the resonant cross section, which is well above both the CLEO data
(\ref{data}) on the $D {\bar D}$ cross section and the average\cite{rosner3} for
the total resonant cross section. Thus it looks reasonable to not be bound by a
particular value of the mixing angle $\theta$ derived from the current data on
the $\ell^+ \ell^-$ decay widths, and possibly to look for alternative hints at
the value of this parameter.

In the picture discussed here a mixing with the $\ppp$, which has a relatively
large annihilation rate into light hadrons due to the four-quark component as
described by eq.(\ref{decomp}), should contribute to the similar decay rate of
the $\pp$ resonance. This extra contribution adds incoherently to the standard
annihilation rate originating in the colorless $c {\bar c}$ pair annihilation
into three gluons. The measured branching fraction in eq.(\ref{datc})
corresponds in absolute terms to the partial width $\Gamma(\pp \to {\rm light~
hadrons}) = (47.5 \pm 8.6)\,$KeV, while a scaling of the $J/\psi$ width
proportionally to the $\ell^+ \ell^-$ widths as shown in eq.(\ref{ppjpth}) would
yield about 30 KeV. Given the uncertainty in the experimental data, and the
possible contribution of the previously discussed non-perturbative effects in
the three gluon annihilation, a choice of 20 KeV as a `representative' value for
the excess in the direct decay width of the $\pp$ resonance does not look
unreasonable. The extra decay width in the mixing scheme is related to the
(small) mixing angle $\theta$ as
\beq
\Delta  \Gamma(\pp \to {\rm light~ hadrons}) = \theta^2 \, \Gamma[\ppp \to {\rm
light~ hadrons}]~,
\label{glh}
\eeq
so that for an $O(10\%)$ branching fraction of the $\ppp$ non-$D {\bar D}$
decays one approximately estimates $\theta \approx 0.1$. It can be noticed that
the estimate\cite{rosner} $\theta = 12^o \approx 0.2$ would result in a
substantially larger, beyond allowed by the experiment, extra decay rate of
$\pp$, provided that the non-$D {\bar D}$ branching fraction for the $\ppp$ is
indeed around 10\%.

Due to the mixing a presence of an $I=1$ isotopic component in $\ppp$ should
result in a small isovector contribution in the wave function of the $\pp$
resonance. For the reasons of the $G$ parity this part does not affect the
isotopic relation between the decays $\pp \to \pi^+ \pi^- \, J/\psi$ and $\pp
\to \pi^0 \pi^0 \, J/\psi$, which relation has recently been successfully
verified by CLEO\cite{cleo2}. However the isovector part gives an extra
contribution to the amplitude of the decay $\pp \to \pi^0 \, J/\psi$. This extra
contribution is necessarily coherent with the one derived in the standard
approach\cite{is}. Indeed, there is only one partial wave ($P$ wave) allowed in
this process, and any contributions are bound to be relatively real by the
absence of essential on-shell intermediate states. Unlike in the standard
approach for a pure $c {\bar c}$ charmonium, for  the $I=1$ four-quark state,
assumed in this discussion, the transition amounts to a rearrangement of quarks
with the $c {\bar c}$ materializing as the $J/\psi$ and the light quarks as the
neutral pion. Therefore one can reasonably assume that even a small admixture of
such four-quark component can produce a sizeable effect in the amplitude. The
discrepancy between the data (\ref{datd}) and the theoretical formula can be
explained by an extra contribution to the decay amplitude amounting to about one
half of the standard one and having the same sign. In other words, the
additional contribution should amount to about one third of the amplitude
corresponding to the observed decay rate in eq.(\ref{datd}).

\section{The decays $\ppp \to \pi^0 \, J/\psi$ and $\ppp \to \eta \, J/\psi$}
The explanation of the somewhat enhanced rate of the hadronic transition $\pp
\to \pi^0 \, J/\psi$ as being due to a small admixture of the $I=1$ four-quark
component of $\ppp$ can be turned around to predict the rate of the yet
unobserved decay $\ppp  \to \pi^0 \, J/\psi$, since in this decay the $I=1$ part
of the four-quark component should dominate the amplitude:
\beq
\Gamma[\ppp  \to \pi^0 \, J/\psi] \approx {(1 \div 1.5) \over \theta^2} \left (
{1 \over 3} \right )^2 \, \Gamma(\pp \to \pi^0 \, J/\psi) \approx (4 \div 6) \,
\left ( {0.1 \over \theta} \right )^2 \, {\rm KeV}~,
\label{ppppijp}
\eeq
where the factor (1/3) corresponds to the assumed fraction of the actual
amplitude of $\pp \to \pi^0 \, J/\psi$ being due to the mixing, and the range
$(1 \div 1.5)$ refers to whether the rescaled $P$ wave kinematical factor
$p_\pi^3$ is included in the estimate or not. In terms of the branching fraction
${\cal B}[\ppp  \to \pi^0 \, J/\psi]$ the estimate (\ref{ppppijp}) corresponds
to about $2 \times 10^{-4}$ (at $\theta \approx 0.1$). While certainly small,
such branching fraction does not look totally unrealistic to be measured given
the total number of events in the CLEO-c sample.

The assumed $I=0$ four-quark component naturally enters with a larger weight
than $I=1$ and its extra contribution to the amplitude of the decay $\pp \to
\eta \, J/\psi$  could potentially jeopardize the agreement between the
experiment and the theoretical prediction\cite{vz} for the rate of this decay 
(relative to $\pp \to \pi \pi \, J/\psi$). In order to analyze this situation,
we write the SU(3) relation between the extra contributions to the amplitudes in
terms of the coefficients $b_0$ and $b_1$ in eq.(\ref{decomp})
\beq
{\Delta A(\pp \to \eta \, J/\psi) \over \Delta A(\pp \to \pi^0 \, J/\psi)}={b_0
\over \sqrt{3} \, b_1}~,
\label{rdel}
\eeq
while experimentally the ratio of the absolute values of the amplitudes, after
taking into account the kinematical factor $p_{\pi,\eta}^3$, is equal to
approximately 22, and, as previously suggested, the extra contribution  amounts
to about one third of the actual amplitude of the decay $\pp \to \pi^0 \,
J/\psi$. As estimated originally\cite{vz}, the uncertainty in the theoretical
prediction of the amplitude of the decay $\pp \to \eta \, J/\psi$ is about 20\%.
Thus the extra contribution would not exceed this uncertainty as long as the
ratio $|b_0/b_1|$ is less than approximately 23, i.e. as long as $|b_1|\, \gsim
\, 0.043 \, |b_0|$. This constraint is fully compatible with the estimates
presented here, so that the agreement between the standard theory and the
experiment for the decay  $\pp \to \eta \, J/\psi$ is safe from being
invalidated by the extra contribution beyond its intrinsic uncertainty.

The $I=0$ part of the four-quark component can however give rise to a
realistically measurable rate of the decay $\ppp \to \eta \, J/\psi$. Indeed,
applying the same relation as in eq.(\ref{rdel}) to the decay amplitudes of the
$\ppp$ one estimates
\beq
{\Gamma[\ppp \to \eta \, J/\psi] \over \Gamma[\ppp \to \pi^0 \, J/\psi]}={1
\over 3}\, \left | {b_0 \over b_1} \right |^2 \, {p_\eta^3 \over p_\pi^3}~.
\label{repi}
\eeq
Assuming, as previously, $|b_0|^2 \sim 0.1$ and $|b_1|^2 \sim (0.03)^2 \approx
10^{-3}$, one finds ${\cal B}[\ppp \to \eta \, J/\psi] \sim  7 \, {\cal B}[\ppp
\to \pi^0 \, J/\psi] \sim 0.14 \%  \, (0.1/\theta)^2$. On the other hand, if the
ratio $|b_0/b_1|$ is assumed to be at the discussed upper limit allowed by the
uncertainty in the amplitude of the decay $\pp \to \eta \, J/\psi$, the estimate
for the rate of the $\ppp$ decay substantially increases:
${\cal B}[\ppp \to \eta \, J/\psi] \approx   36 \, {\cal B}[\ppp \to \pi^0 \,
J/\psi] \approx 0.7 \% \, (0.1/\theta)^2$.

\section{Summary and discussion}
It summary. It is suggested here that the wave function of the resonance $\ppp$
contains a substantial four-quark component of the type $c {\bar c} u {\bar u}$
and $c {\bar c} d {\bar d}$ which component is not quite symmetric between the
extra $u$ and $d$ quarks, resulting in a presence of an isovector $I=1$ part.
Moreover, a small fraction of this four-quark state also enters the wave
function of the $\pp$ resonance due to a small $\ppp - \pp$ mixing. As shown,
this assumption allows to explain the possible discrepancies between the
theoretical expectations based on the picture of $\ppp$ and $\pp$ being pure $c
{\bar c}$ states and the recent experimental data. In particular the charmed
quarks inside the four-quark state are allowed to annihilate into light hadrons
at a much higher rate than from a colorless $1^{--}$ state, which results in a
possible fraction of direct decays of the $\ppp$ into light hadrons. The $I=1$
four-quark component of $\ppp$ should create a bias in the coupling of this
resonance to $D^+ D^-$ and $D^0 {\bar D}^0$ which can explain the observed
charged to neutral yield ratio (\ref{datb}) at the resonance peak. The
additional contribution to the direct annihilation rate feeds down to the $\pp$
resonance through the $\ppp - \pp$ mixing and eliminates the discrepancy between
the central value of the measured branching fraction ${\cal B}(\pp \to {\rm
light~ hadrons})$  in eq.(\ref{datc}) and the standard scaling from $J/\psi$ of
the direct annihilation rate proportionally to the $\ell^+ \ell^-$ decay width.
Finally, the $I=1$ part of the four-quark component appearing in $\pp$ as a
result of the mixing can explain the excess of the measured rate of the decay
$\pp \to \pi^0 \, J/\psi$ (eq.(\ref{datc})) over the long-standing theoretical
predictions.
The suggested four-quark admixture in the $\ppp$ resonance should result in yet
unobserved decays $\ppp \to \pi^0 \, J/\psi$ and $\ppp \to \eta \, J/\psi$. The
estimates for the rates of these decays are still quite uncertain, but the
`typical' expected values for their branching fractions:  ${\cal B}[\ppp \to
\pi^0 \, J/\psi] \sim 2 \times 10^{-4}$ and ${\cal B}[\ppp \to \eta \, J/\psi]
\sim 0.15 \%$, illustrate that an experimental search for these processes is
quite feasible.

At this point no dynamical scheme is offered for a more detailed consideration
of the four-quark component of the $\ppp$. Generally it may be related to the
strong coupling of the resonance to the $D {\bar D}$ channel, but may also be
affected by `molecular charmonium'\cite{ov} effects. I believe that a more
specific scheme than presented here would receive a strong boost from improved
experimental data, especially if the decays $\ppp \to \pi^0 \, J/\psi$ and $\ppp
\to \eta \, J/\psi$ are found at or around the level indicated above. It should
be also mentioned that in some of the discussed cases the deviation of the
experimental data from the theoretical expectations just barely exceeds $2
\sigma$. In particular this is the case for the direct annihilation rate of the
$\pp$ in eq.(\ref{datc}), and also for the difference between the total and the
$D {\bar D}$ resonant cross section at the $\ppp$ resonance, where the total
cross section is not yet available from the CLEO data, and one has to
resort\cite{rosner3} to averaging the results of the previous experiments. The
confusion around the data is further enhanced by that the available experimental
numbers do not appear to agree well with each other in view of the unitarity
formula in eq.(\ref{smax}) and thus at least some of the data will have to
change. Thus for a general understanding of the properties of the charmonium
resonances and for a better assessment of the status of a mixing scheme along
the lines discussed here it is quite important that sufficiently accurate and
reliable data become available at least for such basic characteristics of the
$\ppp$ resonance as its total width and the total resonant $e^+e^-$ cross
section.

\section*{Acknowledgements}
I thank Dan Cronin-Hennessy, Yuichi Kubota, and Misha Shifman for useful
discussions of the experimental and theoretical topics related to this paper.

This work is supported in part by the DOE grant DE-FG02-94ER40823.

\end{document}